\documentclass[%
 reprint,
 amsmath,amssymb,
 aps,
]{revtex4-1}

\usepackage{graphicx}
\usepackage{dcolumn}
\usepackage{bm}

\usepackage{caption}
\usepackage{subcaption}
\usepackage{mhchem}

\newcommand{\tr}{\mbox{tr}}

\newcommand{\unit}[1]{\ensuremath{\, \mathrm{#1}}}

\begin{document}

\preprint{APS/123-QED}

    \title{Quantum Melting in a Polariton Lattice}
    \date{\today}
    \author{Alexander Edelman}
    \affiliation{James Franck Institute, 929 East 57th Street, Chicago, Illinois 60637, USA}
    \affiliation{Department of Physics, The University of Chicago, 5720 South Ellis Avenue, Chicago, Illinois 60637, USA}
    \author{Peter B. Littlewood}
    \affiliation{James Franck Institute, 929 East 57th Street, Chicago, Illinois 60637, USA}
    \affiliation{Department of Physics, The University of Chicago, 5720 South Ellis Avenue, Chicago, Illinois 60637, USA}
    
    \begin{abstract}
        Inspired by the recent experimental observation of strongly coupled polaritons in a Moiré heterobilayer, we study a model of dipole-interacting excitons localized on sites of a lattice and coupled to planar cavity photons. We calculate the phase diagram of this system by computing fluctuations around the mean field and determining the stability of the excitation spectrum. We find that the transition from the normal state to a polariton condensate is intermediated by a series of ordered states at partial fillings of the exciton lattice, stabilized by the exciton interactions. In particular we predict a supersolid phase in which a polariton condensate coexists with spatial order. 
    \end{abstract}
    
    \maketitle
    
        Polaritons are bosonic quasiparticles that arise when the coupling in a systen of light and matter is sufficiently strong to hybridize the two components. Since the observation of Bose-Einstein condensation in a realization with quantum well excitons coupled to the cavity photons in a distributed Bragg reflector, they have been widely studied as both a platform for exploring condensation phenomena (including at room temperature due to the low effective mass of the cavity photon), as well as a practical photonic nonlinearity for optical computation and quantum simulation \cite{Kasprzak_Richard_Kundermann_Baas_Jeambrun_Keeling_Marchetti_2006, Kena-Cohen_Forrest_2010, Su_Ghosh_Wang_Liu_Diederichs_Liew_Xiong_2020, Schneider_Winkler_Fraser_Kamp_Yamamoto_Ostrovskaya_Hofling_2016, Suarez-Forero_Riminucci_Ardizzone_Gianfrate_Todisco_Giorgi_Ballarini_Gigli_Baldwin_Pfeiffer_2021}. A recent thrust of experimental efforts has been toward establishing the transition metal dichalcogenides (TMDs) as a polaritonic system\cite{Hu_Fei_2020}. Besides their attractive optical properties, the TMDs naturally exhibit strong interactions that can already realize exotic electronic phases of matter such as Wigner crystals\cite{Zhou_Sung_Brutschea_Esterlis_Wang_Scuri_Gelly_Heo_Taniguchi_Watanabe_2021,Smolenski_Dolgirev_Kuhlenkamp_Popert_Shimazaki_Back_Lu_Kroner_Watanabe_Taniguchi_2021}. Stacking multiple layers of TMDs that have been twisted relative to each other generates a Moiré superlattice potential and flattens the low-energy band structure to further exacerbate interaction effects, leading to the observation of more dilute electronic crystals in which charge carriers localized on Moiré lattice sites organize into spatially ordered states\cite{Miao_Wang_Huang_Chen_Lian_Wang_Blei_Taniguchi_Watanabe_Tongay_2021,Xu_Liu_Rhodes_Watanabe_Taniguchi_Hone_Elser_Mak_Shan_2020,Huang_Wang_Miao_Wang_Li_Lian_Taniguchi_Watanabe_Okamoto_Xiao_2021,Jin_Tao_Li_Xu_Tang_Zhu_Liu_Watanabe_Taniguchi_Hone_2021}. Excitons in such structures exhibit strong dipolar interactions and can localize, and a recent experiment has realized strong-coupling polaritons in a Moiré superlattice in a cavity and found evidence of strong nonlinearities associated with the localized excitons saturating at a density of one per Moiré lattice site\cite{Li_Lu_Dubey_Devenica_Srivastava_2020, Zhang_Wu_Hou_Zhang_Chou_Watanabe_Taniguchi_Forrest_Deng_2021}. 

\begin{figure}
            \includegraphics[width=\columnwidth]{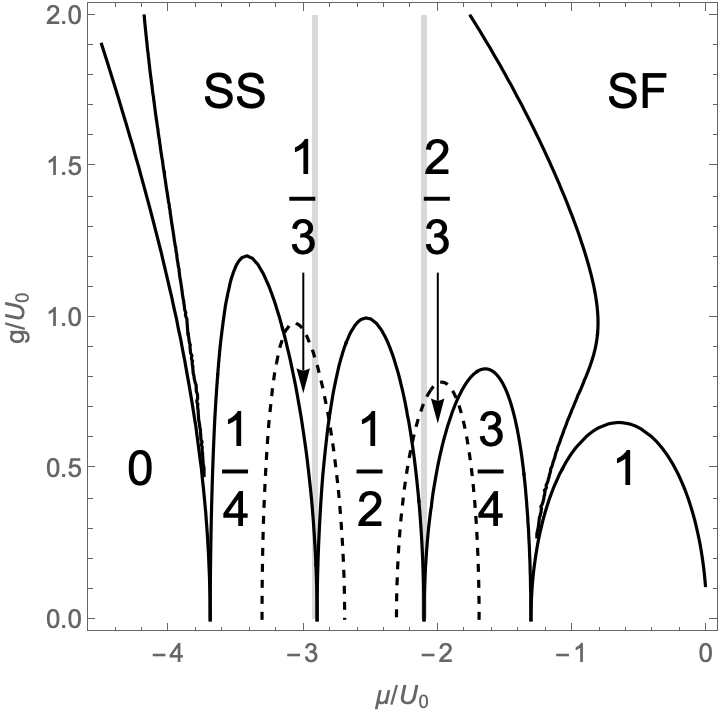}
            \caption{Phase diagram of polaritons on a Moiré lattice. SS: supersolid; SF: superfluid; other phases are labeled by exciton filling. Gray regions are those where $p/3$ phases minimize the energy at $g=0$.}
            \label{fig:phasdiag}
        \end{figure}

        In this work we study a model of polaritons formed from excitons on a lattice coupled to a planar cavity, taking into account possible strong exciton-exciton interactions. We study the case where the lattice spacing is small compared to the photon wavelength and blockade effects restrict occupancy to one exciton per site; this should be contrasted with previous studies of polariton systems with larger-scale spatial order that can be described in terms of the modulation of some continuous density, leading to polariton band structures\cite{Winkler_Fischer_Schade_Amthor_Dall_Gessler_Emmerling_Ostrovskaya_Kamp_Schneider_2015,Lai_Kim_Utsunomiya_Roumpos_Deng_Fraser_Byrnes_Recher_Kumada_Fujisawa_2007,Pickup_Sigurdsson_Ruostekoski_Lagoudakis_2020}. Our main result is shown in Figure~\ref{fig:phasdiag}. We display the phase diagram of the system as a function of chemical potential $\mu$ and light-natter interaction strength $g$, both normalized by the dipolar interaction strength $U$. We find that, as in prior work, the system transitions from an insulating phase to a polariton superfluid as $\mu$ and $g$ are increased. The transition proceeds, however, through a number of intermediate states: at small $g$, a series of correlated excitonic insulators at partial filling of the superlattice, and at larger $g$ a supersolid where a finite occupancy of the photon field coexists with spatial ordering of the matter component.

    \paragraph*{Model}
        The Hamiltonian is $H=H_0 +H_\text{int}$, with
        \begin{equation}
            H_0 = \sum_{\mathbf{q}}\psi^\dagger_{\mathbf{q}}\omega_{\mathbf{q}}\psi_{\mathbf{q}} +\epsilon\sum_j\sigma^z_j +g\sum_{j \mathbf{q}}(e^{i \mathbf{q}\cdot \mathbf{r}_j}\psi^\dagger_{\mathbf{q}}\sigma^-_j +\text{h.c.})
            \label{eq:H0}
        \end{equation}
        and $H_\text{int} = \sum_{jj^\prime}U(|\mathbf{r}_j-\mathbf{r}_{j^\prime}|)\mathcal{P}^\uparrow_j\mathcal{P}^\uparrow_{j^\prime}$. $H_0$ is an extended Dicke model like that studied by Keeling et al\cite{Keeling_Eastham_Szymanska_Littlewood_2004}, where $\psi_\mathbf{q}$ annihilates a photon of in-plane momentum $\mathbf{q}$ and the presence or absence of an exciton on site $j$ with on-site energy $\epsilon$ is represented as the up or down eigenstate, respectively, of the spin operator $\sigma_j^z$. In what follows we will use the language of spin flips interchagneably with exciton occupancy. The in-plane photon dispersion is $\omega_\mathbf{q} = \omega_0 +\mathbf{q}^2/2m^*$ where the fundamental frequency $\omega_0$ and the effective mass $m^*$ are set by the geometry of the cavity. The light-matter term describes an exciton created (annihilated) by absorption (emission) of a photon at site $j$, and the dipolar coupling strength $g$ can be inferred from the measured Rabi splitting $\Omega = g\sqrt{n}$ where $n$ is the density of lattice sites. $\mathcal{P}_j^\uparrow$ projects onto the spin-up state at site $j$, so that $H_\text{int}$ describes a pairwise interaction $U$ between occupied sites.

        Working in the grand canonical ensemble, we use the method of Fedotov and Popov\cite{Popov_Fedotov_1988} to represent the spins by ``semions" subject to an on-site occupancy constraint, represent the system in Fourier space, and use standard techniques to integrate out these fermions after decoupling the interaction in the density channel by a bosonic field $\phi$\cite{Altland_Simons_2009}. We obtain the effective action

        \begin{equation}
            \begin{split}
                S_{\text{eff}}[\psi,\phi] =& \sum_{q}\psi^\dagger_{q}(-i\omega_n +\tilde\omega_q)\psi_{q} -\sum_{q} \phi_{q} U(q) \phi_{-q} \\
                &-\tr\ln\mathcal{M} \\
                \mathcal{M}^{-1}_{kk^\prime} =& \left(\begin{array}{cc}
                    -i\nu_m -\xi_k & 0 \\
                    0 & -i\nu_m +\xi_k
                \end{array}\right)\delta_{kk^\prime} \\
                &+ \sum_q \left(\begin{array}{cc}
                    -|U(q)|\phi_{-q} & g\psi^\dagger_{-q} \\
                    g\psi_q & |U(q)|\phi_q
                \end{array}\right)\delta_{k,k^\prime-q}.
            \end{split}
            \label{eq:seff}
        \end{equation}
    
        Here $\omega_n$ ($\nu_m$) is a bosonic (semionic) Matsubara frequency, $\tilde\omega_q = \omega_q -\mu$, $\phi_q = \langle \sum_j e^{i\mathbf{q}\cdot\mathbf{r}_j}\sigma^z_j\rangle$, $\xi_k = (\epsilon - \mu)/2$, $\mu$ is the chemical potential, and $\mathcal{M}$ lives in the space of up and down spins $\times$ 4-momenta $k=(m,\mathbf{k})$.

        We will consider a number of mean-field Ans{\"a}tze for $\langle \psi \rangle$ and $\langle \phi \rangle$. Because the Moiré lattice spacing is much smaller than the photon wavelength, for energetic reasons we only consider the possibility of a spatially uniform photon condensate, $\langle \psi_0 \rangle = \lambda$. In each case we will minimize the free energy, then expand in fluctuations $\delta\psi$ and $\delta\phi$. The general form of the fluctuation contribution to the free energy is $\delta f = \sum_q \mathbf{\Delta}_q\cdot\mathbf{Q}_q\cdot\mathbf{\Delta}_{-q}^\dagger$ with $\mathbf{\Delta}_q = \left(\begin{array}{ccc}\delta\psi_q^\dagger & \delta\psi_{-q} & \delta\phi_q \end{array}\right)$ and
        \begin{equation}
           \mathbf{Q}_q = \left(\begin{array}{ccc}
    			-i\omega_q +\tilde\omega_q + K_1 & K_2\lambda^2 & K_3\lambda\\
    			K_2^*(\lambda^*)^2 & i\omega_q +\tilde\omega_q + K_1^* & K_4\lambda^* \\
    			K_3^*\lambda^\dagger & K_4^*\lambda & -U(q)+\Pi|\lambda|^2 
    		\end{array}\right).
            \label{eq:flucts}
        \end{equation}
        The coefficients are calculated in the SI. Notably, in the absence of a condensate amplitude $\lambda$ the propagating and counter-propagating photon fluctuations are decoupled from each other and from the density fluctuations. Physically, this reflects the absence of a kinetic energy term in the Hamiltonian for the excitons, owing to their orders of magnitude larger effective mass compared to the photon. The spectrum $S(\omega,q)$ is given by the poles of the fluctuation Green's function and is determined by analytic continuation $i\omega_q \to \omega$ and solving $\det\mathbf{Q}_q(\omega) = 0$. The phase diagram is constructed by finding the lines of instability of each phase, diagnosed by a softening of the spectrum $S(\omega \to 0, q) \to 0$. Although our formalism is equipped to handle thermal effects, here we will discuss $T=0$ results.

    \paragraph*{Exciton Chain, Normal State}
        \begin{figure*}
            \begin{subfigure}{.32\linewidth}
                \includegraphics[width=\linewidth]{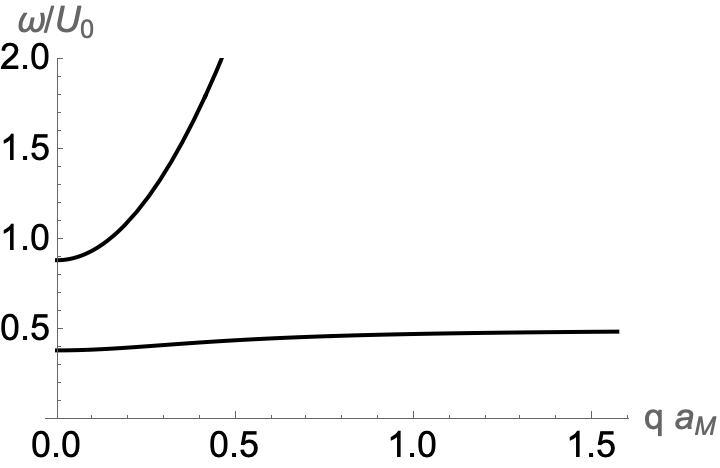}  
                \caption{}
                \label{fig:phi0spec}
            \end{subfigure}
            \begin{subfigure}{.32\linewidth}
                \includegraphics[width=\linewidth]{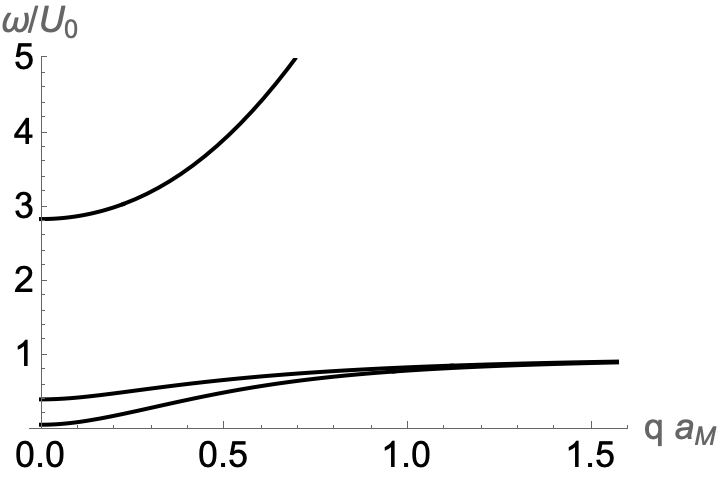}  
                \caption{}
                \label{fig:phiQspec1}
            \end{subfigure}
            \begin{subfigure}{.32\linewidth}
                \includegraphics[width=\linewidth]{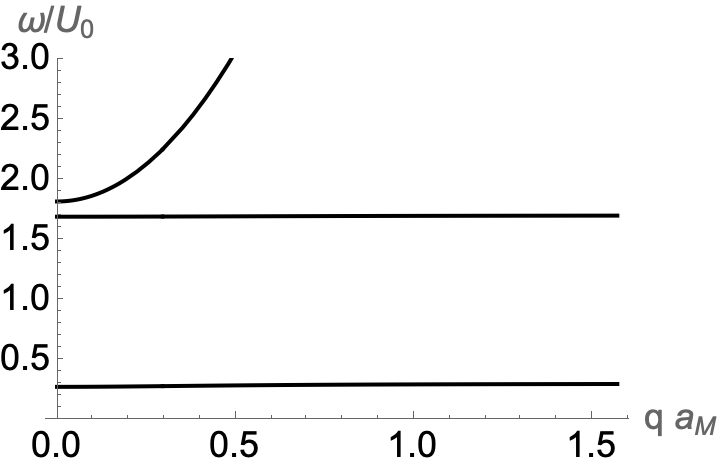}
                \caption{}
                \label{fig:phiQspec2}
            \end{subfigure}
            \caption{Excitation spectra of normal states with $\Delta^*=2.5$, $m^*=.1$. a) Fully polarized state at $g^*=.4$, $\mu^*=-1$. b) Staggered state at $g^*=1.1$, $\mu^*=-2.5$. c) Staggered state at $g^*=.2$, $\mu^*=-1.8$}
            \label{fig:spectra}
        \end{figure*}
        To illustrate the essential physics we start by considering a one-dimensional chain of exciton sites with repulsive nearest-neighbor interactions only, $U(q) = U_0\cos(qa_M)$ where $a_M$ is the lattice spacing. In what follows we will nondimensionalize by $U_0$ and $a_M$, so that the model is controlled by the effective detuning $\Delta^* = (\omega_0 -\epsilon)/U_0$, Rabi splitting $g^* = g\sqrt{n}/U_0$, chemical potential $\mu^*=\mu/U_0$, and mass $m^* = U_0 m /\hbar^2 \sqrt{n}$ where $n=1/a_M^2$ is the density of sites.
        
\begin{figure}
            \includegraphics[width=\columnwidth]{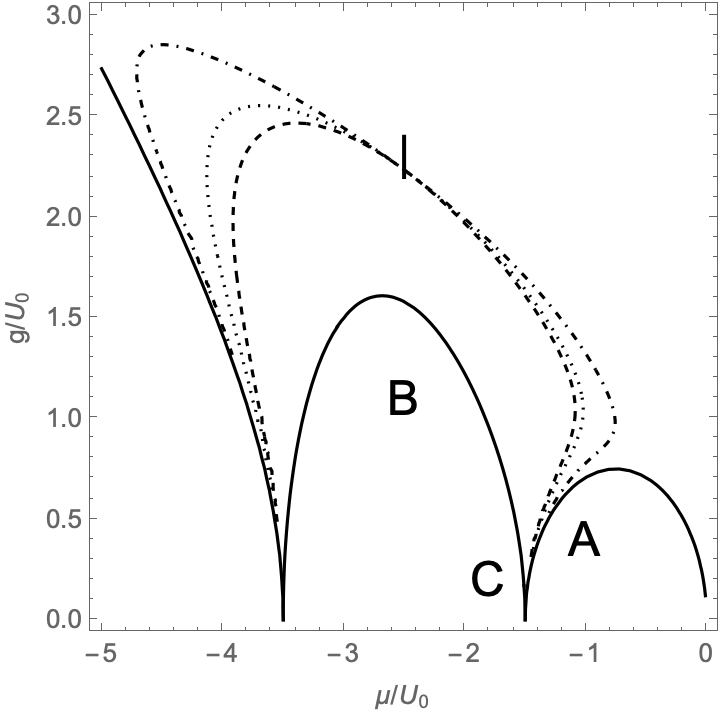}
            \caption{Phase diagram of the exciton chain. Phase boundaries are determined by stability of the spectra. Solid: normal-to-condensate boundaries. Superfluid-to-supersolid boundaries are computed for $m^*=10^{-6}$ (dashed), $m^*=.1$ (dotted), $m^*=1$ (dot-dashed). Labeled points correspond to parameters in Figure~\ref{fig:spectra} and vertical line cut corresponds to Figure~\ref{fig:instability}.}
            \label{fig:mstar}
        \end{figure}

        We begin by considering possible normal states, i.e. excitonic insulators with $\lambda = 0$. The simplest uniform exciton population minimizes the free energy in two fully polarized states, $\langle \phi_0 \rangle = \pm \frac{1}{2}$. The excitations, shown in Figure~\ref{fig:phi0spec}, are the usual upper and lower polaritons. Note that we have plotted  When $g$ is of order the energy of a spin flip, the system suffers a $q=0$ instability to condensate formation. The instability lines in Figure~\ref{fig:mstar} therefore form a ``Mott lobe'' structure with the (zero-temperature) condensate persisting down to $g=0$ when the chemical potential is tuned so that the system is indifferent between normal states. The phase diagram is shown with the exciton detuned $\Delta^*=2.5$ below the photon energy; above $\mu=0$ photons proliferate and the system always condenses.

        The other normal state that this simple choice of interaction can stabilize is staggered configuration of alternating exciton occupancy, which can conveniently described as a density wave at $q=\pi a_M$. The free energy is again minimized at full polarization of the state, $\langle \phi_\pi \rangle = \frac{1}{2}$, and the system is once again unstable to condensation, producing a third Mott lobe that completes the $g=0$ phase diagram. Figures~\ref{fig:phiQspec1} and \ref{fig:phiQspec2} respectively show spectra at large $g$ near the center of the lobe, and at small $g$ near its edge. In this phase the spin flip excitations on each sublattice are nondegenerate, leading to two lower polaritons whose splitting is controlled by the chemical potential. The photon couples more strongly to the lower of these, leading eventually to the $q=0$ instability to condensation. As the chemical potential is changed in the $g\to0$ limit, the transition to a new ground state is signaled by a softening of the entire lower branch.

    \paragraph*{Instability of the Condensate}
        \begin{figure}
            \includegraphics[width=\linewidth]{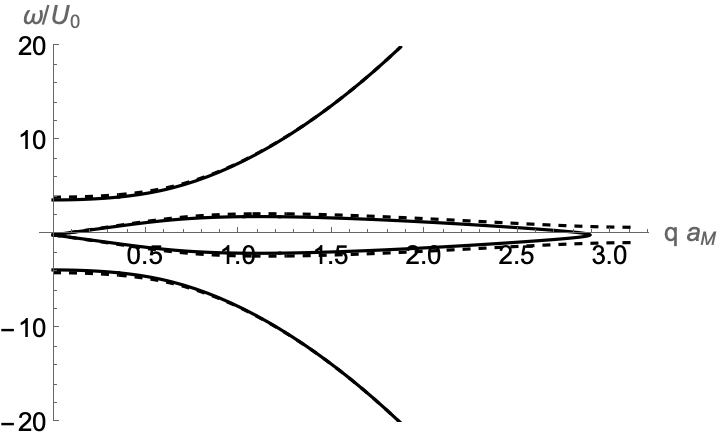}  
            \caption{Excitation spectra of the uniform condensed state at $\mu^*=-2.5$ with a stable lower polariton at $g^*=2.4$ (dashed) and unstable at $g^*=2.2$ (solid).}
            \label{fig:instability}
        \end{figure}
        In the homogeneous condensed phase both $\lambda = \langle \psi_0 \rangle$ and $\langle \phi_0 \rangle$ take on continuous values. At the mean-field level $\lambda$ acts like a transverse magnetic field for the spins, lowering the system energy by their incomplete polarization. The off-diagonal elements of (\ref{eq:flucts}) become non-zero, coupling propagating and counter-propagating photon fluctuations both to each other and density fluctuations. Consequently the lower polaritons in the $q\to0$ limit become the sound modes of the condensate while the upper polaritons become the amplitude modes, and negative energy (i.e. below the chemical potential) partners of these excitations appear, as illustrated near $q=0$ in Figure~\ref{fig:instability}. Crossing the phase boundary to the fully polarized normal states causes the sound mode to become diffusive, and this instability line also saturates the mean-field saddle condition $\delta S_\text{eff}[\psi_0]/\delta \psi_0 = 0$ with $\psi_0 = 0$. 

        As $g$ is lowered at intermediate chemical potentials, however, the lower polariton softens at $q=\pi$ and becomes unstable well before reaching the staggered normal state. Figure~\ref{fig:instability} shows spectra on both sides of the instability, along the cut plotted on Figure~\ref{fig:mstar}. The stability criterion reads

        \begin{equation}
            \label{eq:stability}
            U(q) = \frac{-\tilde\omega_q E^2 -\mu\bar\xi^2}{2\tilde\omega_q|\psi_0|^2\mu}
        \end{equation}

        where $E^2 = g^2|\psi_0|^2 + \bar\xi^2$ and $\bar\xi = \xi+U_0\phi_0$ is the renormalized on-site energy. The instabilities of the uniform condensate are plotted in Figure~\ref{fig:mstar} for different values of $m^*$. As $m^* \to 0$ the instability line approaches $U(q) = -E^2/2\mu|\psi_0|^2$, while for larger effective mass the unstable region occupies a greater part of the phase diagram and exhibits considerable reentrance.

        The region of the phase diagram between the instability lines of the staggered phase and the uniform condensate is a supersolid, in the sense that the condensate order parameter $\langle \psi_0 \rangle$ is nonzero while the exciton condensate exhibits spatial order. We have verified through Monte Carlo simulations of the spin system coupled to the photon mean field that this phase is energetically favored, and that the mean-field phase boundary coincides with the $m^*=0$ limit of (\ref{eq:stability}). Physically, the free energy in the presence of a condensate is minimized by incomplete polarization of the spins, which is always disfavored by the excitonic component alone. At intermediate $g$ when the photon order is relatively weak, the total density can be better accommodated by only partially filling the excitons - even at values of the chemical potential that favor full polarization in the normal state. In the $m^*=0$ limit, the photon fluctuations are instantaneous. The greater extent of the supersolid phase with finite $m^*$ reflects the lower superfluid stiffness (as measured by the sound velocity) and hence greater propensity to destabilize.

    \paragraph*{Moiré Lattice}
        We can generalize these arguments to study the system on the triangular Moiré lattice seen in experiment. We include now nearest- ane next-nearest-neighbor interactions and assume that they are dipolar with $U(r)\sim 1/r^3$ so that the next-nearest $U_1 = U_0/3\sqrt{3}$, although our results depend only quantitatively on this choice. Finding the normal state phase diagram at $g=0$ is isomorphic to the lattice gas model studied by Kaburagi and Kanamori\cite{Kaburagi_Kanamori_1974}, and our chosen interaction narrowly satisfies the condition $U_0 > 5 U_1 > 0$ to stabilize phases of $1/4$, $1/3$, and $1/2$ filling (and their complements). The half-filled phases are stripes commensurate with the quarter-filled phases which introduce a further staggered order within a stripe, while the third-filled phases have different symmetries.
        
        The density-wave order parameter introduced to describe the staggered chain is inconvenient for this richer set of configurations because higher harmonics would be necessary to produce the appropriate occupancy on each lattice site. Instead we introduce an appropriate number of sublattices for each phase and define separate spin variables on each sublattice. Thus, with $N$ sublattices, $\mathcal{M}$ is enlarged to $2N\times 2N$, $\phi_q$ is generalized to $\mathbf{\Phi}_q = \left(\begin{array}{ccc} \phi_q^1 & \cdots & \phi_q^N \end{array}\right)$ and $U$ in (\ref{eq:seff}) must be replaced by a matrix describing the interactions between each sublattice. The number of density fluctuations is likewise enlarged but (\ref{eq:flucts}) retains its general structure, and in particular the property that the photon and density sectors couple only in the presence of a condensate. The normal state spectrum with up to $N$ nondegenerate lower polaritons therefore determined by the contents of $K_1$ and $K_1^*$, and the long-wavelength softening of this spectrum determines the stability of the normal state. Meanwhile the extension of (\ref{eq:stability}) to the appropriate $U(q)$ is benign.

        Together these considerations determine the phase diagram in Figure~\ref{fig:phasdiag}. For clarity the states commensurate with $p/4$ filling have had their phase boundaries drawn as solid lines, while the $p/3$ are bounded by dashed lines. The gray regions depict the chemical potential ranges where the $p/3$ fillings are energetically favored at $g=0$. The transition between incommensurate fillings will be first-order in both normal and supersolid states. Barring a change in symmetry, however, the supersolid is able to continuously vary the exciton density on each sublattice.

        The inclusion of longer-range interactions will stabilize further normal states at intermediate fillings, eventually forming a two-dimensional analog of the devil's staircase\cite{Dublenych_2009}, although given the rapid fall-off of the dipolar interaction and the already narrow range of chemical potentials that favors the $p/3$ phases, we expect these to be difficult to observe. As seen already with second-neighbor interactions in Figure~\ref{fig:phasdiag}, phases with larger regions of stability at $g=0$ penetrate deeper into the supersolid regime (up to the overall effect of larger chemical potential favoring condensation), and we expect the supersolid to accordingly inherit the symmetry of these phases. 

    \paragraph*{Experimental Signatures}
        The clearest spectral signature of a spatially ordered phase in our model is the splitting by the interaction of the lower polariton into distinct modes on each sublattice. (We note that the middle polaritons observed in \cite{Lai_Kim_Utsunomiya_Roumpos_Deng_Fraser_Byrnes_Recher_Kumada_Fujisawa_2007} come from the presence of both inter- and intra-layer Moiré excitons, a distinct effect that we have not considered here.) Spectra are computed relative to the chemical potential. Experimentally the density is controlled by pumping the system, and the blue shift of the resulting emission provides a measure of the chemical potential. At $g=0$, $\mu(\rho)$ (where $\rho=\frac{1}{2} +\phi + |\psi|^2$ is the density) will exhibit a series of jumps as the system transitions between states of different filling. This effect was already predicted in \cite{Eastham_Littlewood_2000} for the transition between homogeneous fully polarized states. At finite $g$, the transitions between normal states and regions of the phase diagram with superfluid order will instead be marked by kinks in $\mu(\rho)$, which will then change continuously in the presence of a superfluid.

        The visibility of these effects is attenuated by noise sources including inhomogeneous broadening of the excitons, thermal effects, and non-equilibrium physics associated with the driven-dissipative nature of the system. The energy scale on which these signatures appear is set by the nearest-neighbor coupling strength, which in the experiments of \cite{Zhang_Wu_Hou_Zhang_Chou_Watanabe_Taniguchi_Forrest_Deng_2021} is $U_0 \sim .5\unit{meV}$, compared to an exciton inhomogeneous broadening of $\sim 8\unit{meV}$ and cavity linewidth of $\sim 3\unit{meV}$, and on the same order as the thermal energy at $5 \unit{K}$. To achieve strong coupling the Rabi splitting must be larger than these, and indeed with $\Omega\sim 10\unit{meV}$ locates the experiment in a region of the phase diagram where only a phase transition from a completely polarized normal state to a uniform condensate is expected. From the measured cavity dispersion we have extracted $m\sim10^{-5}m_e$, or $m^*\sim10^{-6}$, firmly within the mean-field regime. Condensation phenomena are not yet evident in the Moiré polariton system, although they have been achieved in monolayer TMDs\cite{Zhao_Su_Fieramosca_Zhao_Du_Liu_Diederichs_Sanvitto_Liew_Xiong_2021}. 

        In conclusion, we have proposed a simple model that realizes a polariton supersolid phase and may be within reach of present experiments. Our model is closely related to the lattice supersolid phases of hard-core bosons, as realized for instance in an ultracold atomic gas in an optical lattice or helium adsorbed on graphite\cite{Baumann_Guerlin_Brennecke_Esslinger_2010, Choi_Zadorozhko_Choi_Kim_2021}. Our model differs in that what would ordinarily be an off-diagonal order parameter in the matter field is here the photon field, imbued with its own dynamics, leading to a reentrant phase diagram. We expect that this would be substantially enriched as the Moiré lattice spacing is increased to become comparable to the photon wavelength, as on the one hand the importance of the exciton-exciton interaction is enhanced by diluteness, and on the other it becomes energetically possible for the light field to condense in a state with a density modulation commensurate with the cavity. In contrast to polariton band structures in which condensation away from $k=0$ appears as a metastable non-equilibrium effect, the constraint of a single emitter per lattice site would make such supersolids possible in equilibrium. The model is possibly also of relevance to self-organized lattices, for instance due to blockade effects in Rydberg excitons\cite{Kazimierczuk_Frohlich_Scheel_Stolz_Bayer_2014,Bao_Liu_Xue_Zheng_Tao_Wang_Xia_Zhao_Kim_Yang_2019}. We have previously studied a similar model in the context of weak crystallization\cite{Edelman_Littlewood_2015}.
    
    \paragraph*{Acknowledgements}
        We are grateful to R.T. Brierley and R. Hanai for useful discussions.

    \bibliography{supersolid}

\end{document}